\documentclass{article}

\usepackage{arxiv}

\usepackage[utf8]{inputenc} 
\usepackage[T1]{fontenc}    
\usepackage{hyperref}  
\usepackage{graphicx}
\usepackage[export]{adjustbox}
\usepackage{xcolor}
\newcommand\Tstrut{\rule{0pt}{2.1ex}}       
\newcommand\Bstrut{\rule[-0.9ex]{0pt}{0pt}} 
\newcommand{\TBstrut}{\Tstrut\Bstrut} %
\newcommand{\foo}{\hspace{-2.3pt}\textcolor{blue}{$\bullet$} \hspace{5pt}}
\usepackage{subfigure}
\usepackage{paralist}
\usepackage{url}            
\usepackage{booktabs}       
\usepackage{amsfonts}       
\usepackage{nicefrac}       
\usepackage{microtype}      
\usepackage{lipsum}	
\title{Hope Speech Detection: A Computational Analysis of the Voice of Peace}

\author{
  Shriphani Palakodety$^*$\\
  Onai\\
  \texttt{spalakod@onai.com} \\
   \And
Ashiqur R. KhudaBukhsh\thanks{Shriphani Palakodety and Ashiqur R. KhudaBukhsh are equal contribution first authors. Ashiqur R. KhudaBukhsh is the corresponding author.} \\
  Carnegie Mellon University \\
  \texttt{akhudabu@cs.cmu.edu} \\
  \And
Jaime G. Carbonell \\
  Carnegie Mellon University\\
  \texttt{jgc@cs.cmu.edu} \\
}

\begin{document}
\maketitle

\begin{abstract}

The recent Pulwama terror attack (February 14, 2019, Pulwama, Kashmir) triggered a chain of escalating events between India and Pakistan adding another episode to their 70-year-old dispute over Kashmir. The present era of ubiquitious social media has never seen nuclear powers closer to war.
In this paper, we analyze this evolving international crisis via a substantial corpus constructed using comments on YouTube videos (921,235 English comments posted by 392,460 users out of 2.04 million overall comments by 791,289 users on 2,890 videos). Our main contributions in the paper are three-fold. First,  we present an observation that polyglot word-embeddings reveal
precise and accurate language clusters, and subsequently construct a document language-identification technique with negligible annotation requirements. We demonstrate
the viability and utility across a variety of data sets involving several low-resource languages. Second, we present an analysis on temporal trends of pro-peace and pro-war intent observing that when tensions between the two nations were at their peak, pro-peace
intent in the corpus was at its highest point. 
Finally, in the context of heated discussions in a politically tense situation where two nations are at the brink of a full-fledged war, we argue the importance of automatic identification of user-generated web content that can diffuse hostility and address this prediction task, dubbed \emph{hope-speech detection}.
\end{abstract}


\section{Introduction}

\emph{``In peace, sons bury their fathers. In war, fathers bury their sons.''.}\\ -- This comment quoting Herodotus, was automatically discovered by our \emph{hope speech} classifier in our corpus.

On February 14, 2019, a suicide bomber attacked a convoy of vehicles carrying Indian Central Reserve Police Force (CRPF) personnel in Pulwama district, Jammu and Kashmir, resulting in the deaths of 40 CRPF service-personnel and the attacker.
 A Pakistan-based Islamist militant group claimed responsibility, though Pakistan condemned the attack and denied any connection to it. The Pulwama attack triggered a chain of events where each passing day led to an escalation of tensions between India and Pakistan reaching a peak on the $27^{th}$ of February, 2019. With the two nuclear powers coming precariously close to declaring a full-fledged war, the world witnessed a first-of-its-kind specter of war between nuclear adversaries in the modern era of ubiquitous internet, where a unique \emph{war-dialogue} took place between the two nations' civilians on social media.

In this paper, we focus on the \emph{discourse} that took place in comments posted on YouTube - one of the most popular social media platforms in the Indian sub-continent. We collected a comprehensive data set of comments posted in response to YouTube videos on news coverage of relevant incidents by Indian, Pakistani and global media, and analyzed several important aspects of the dialogue between the two conflicting neighbors in relation to this crisis. To the best of our knowledge, ours is the first large-scale analysis of an evolving international crisis between two nuclear adversaries at the brink of a full-fledged war through the lens of social media. India and Pakistan have a long history of political tension that includes four wars and multiple skirmishes resulting in significant military and civilian casualties~\cite{schofield2010kashmir}. Recent scientific analysis projects 100 million deaths in the event of a full-blown war between these two nuclear powers~\cite{toon2019rapidly}. As previously presented in~\cite{zeitzoff2017social}, social media would play an increasingly important role in understanding and analyzing modern conflicts, and we believe that our work would complement the vast literature of quantitative political science on conflict analysis (see, e.g.,~\cite{gochman1983realpolitik}).

\noindent\textbf{Contributions:} Our contributions are the following: 
\begin{compactenum}
    \item \emph{Linguistic:} We present a novel language identification technique that requires minimal human annotation based on the observation that polyglot word-embeddings reveal precise and accurate language clusters. Our technique has applications in analysis of social media content in multilingual
    settings like India, a country with tremendous linguistic diversity (22 major recognized languages) featuring several low-resource languages.
    
    \item \emph{Social:} Through an extensive polarity phrase lexicon, we analyze the temporal trends of pro-war and pro-peace intent and observe that the pro-peace intent reached its peak when the two nations were closest to declaring a full-fledged war.  
    
    \item \emph{Hope speech:} We propose a novel task: \emph{hope-speech detection} to automatically detect web content that may play a positive role in diffusing
      hostility on social media triggered by heightened political tensions during a conflict. Our results indicate that such web content can be automatically identified with considerable accuracy. Solutions to detect hostility-diffusing comments may also find applications in many other contexts. For instance, hostile messages and rumors on platforms like WhatsApp have been used to incite communal violence in the Indian subcontinent in recent times. The severity of the issue prompted the then administration to disable internet access in the regions of unrest to prevent further spread of hateful messages\footnote{\url{https://www.wired.co.uk/article/whatsapp-web-internet-shutdown-india-turn-off}}. Beyond a warlike situation, we expect our work to find application in these and other similar settings.
\end{compactenum}

\section{Background}

\noindent\textbf{A brief history of the conflict:} Kashmir has been a point of contention between India and Pakistan for nearly 70 years. A key factor for continued unrest in South-East Asia, the Kashmir issue has drawn wide attention from the political science community for decades~\cite{malik2002kashmir, schofield2010kashmir, bose2009kashmir,staniland2013kashmir}. The root of this conflict can be traced back to the independence struggle of India and the subsequent partition into India and Pakistan in 1947. Overall, India and Pakistan have gone to full-fledged war four times (1947, 1965, 1971 and 1999) of which, the 1971 war was the goriest (11,000 killed from both sides) which resulted in the largest number of prisoners of war (90,000 POWs) since the Second World War~\cite{ali1983can}. In the four wars, overall, an estimated 27,650 soldiers were killed and thousands wounded. A timeline outlining some of the key events in the bilateral conflict lasting decades is presented in the Appendix.  

\scalebox{1}{

\begin{tabular}{r |@{\foo} p{0.85\textwidth}}
Feb 14th  & A suicide bomber kills 40 CRPF  personnel at Pulwama, India.\\
Feb 15th & A Pakistan-based Islamist militant group, Jaish-e-Mohammad (JEM), claims responsibility. Pakistan condemns the attack and denies any connection to it.\\
Feb 16th & India withdraws ``most favored nation'' status of Pakistan.   \\    
Feb 18th & Nine people, including four Indian soldiers and a policeman are killed in a gun battle in India-controlled Kashmir. \\
Feb 19th & Pakistan Prime Minister Imran Khan offers assistance to investigate the Pulwama attack. India refuses the offer citing previous attacks.\\
Feb 20th & India halts a bus-service between India-controlled Kashmir and Pakistan-controlled Kashmir.\\
Feb 23rd & India begins a two-day crackdown of separatists in Kashmir heightening tensions further.\\
Feb 26th & India claims an airstrike against JEM training base at Balakot and reports a large number of terrorists, trainers and senior commanders have been killed. Pakistan denies any such casualty count.\\
Feb 27th & As an ominous sign of nuclear threat, Pakistan media reports that Imran Khan chaired a meeting of the National Command Authority,
the overseeing body of the country’s nuclear warheads.\\ 
Feb 27th & An Indian Air Force pilot, Abhinandan, is captured by Pakistani armed forces inside Pakistan air space.\\
Feb 28th & Pakistan announces that they will release Abhinandan as a peace gesture.\\
Mar 1st & Pakistan hands over Wing Commander Abhinandan to India at the Wagah border.\\

\end{tabular}
}

\vspace{0.5cm}

\noindent\textbf{Timeline of the most-recent crisis:} We outline some of the key events relevant to the most-recent crisis\footnote{\url{https://www.cnbc.com/2019/03/01/india-pakistan-conflict-timeline.html}} (presented above). We denote five key events: Pulwama terror attack as \texttt{PULWAMA} (Feb 14, 2019), Balakot air strike  claimed by Indian Government as an act of retaliation as \texttt{BALAKOT} (Feb 26, 2019), Indian Air Force (IAF) wing commander Abhinandan's capture by Pakistan (Feb 27, 2019) as \texttt{IAFPILOT-CAPTURE}, Pakistan Government's subsequent announcement of his release as \texttt{IAFPILOT-RELEASE} (Feb 28, 2019), and Pakistan Government's handing over of the captured Indian pilot as \texttt{IAFPILOT-RETURN} (Mar 1, 2019).  

\section{Related Work}
Due to our work's multi-disciplinary nature, throughout the paper, we introduce relevant literature as and when an existing concept is referred. In what follows, we outline a brief description of different lines of research relevant to our paper.

Existing political science literature focusing on Indo-Pak relations and Kashmir \cite{malik2002kashmir, schofield2010kashmir, bose2009kashmir, staniland2013kashmir} has been extended with recent analyses of the Pulwama terror attack~\cite{pandya2019future, feyyaz2019contextualizing} from different viewpoints. Unlike~\cite{pandya2019future, feyyaz2019contextualizing} where the primary focus in on the geopolitical strategic aspects and policy implications of this event, we instead focus on (1) analyzing temporal trends of war and peace intent as observed in social media discussions and (2) a novel task of hostility-diffusing \emph{hope speech detection}.

Recent lines 
of work have explored polyglot word-embeddings' use in a variety of NLP tasks~\cite{mulcaire-etal-2018-polyglot, mulcaire-etal-2019-polyglot, mulcaire2019low} such as parsing, crosslingual transfer etc. In this work, we first show that polyglot embeddings discover language clusters. We subsequently construct a  language identification 
technique that requires minimal supervision and performs well on short social media texts generated in a linguistically diverse region of the world.

In the spirit of encouraging a convivial web-environment, our work of \emph{hope speech detection} is related to hate speech detection~\cite{schmidt2017survey} and early prediction of controversy-causing posts~\cite{hessel-lee-2019-somethings} with a key difference that we aid web-moderation through finding the \emph{good/positive content} while the other two lines help web-moderation through detecting the \emph{bad/negative content}.

\section{Data set: YouTube Comments}

\noindent\textbf{Why YouTube?}  As of April 2019, the platform
drew 265 million monthly active users (225 million on mobile)\footnote{\url{https://www.hindustantimes.com/tech/youtube-now-has-265-million-users-in-india/story-j5njXtLHZCQ0PCwb57s40O.html}} in India accounting for 80\% of the population with internet access\footnote{\url{https://yourstory.com/2018/03/youtube-monthly-user-base-touches-225-million-india-reaches-80-pc-internet-population}}. In Pakistan, 73\% of the population with internet access views YouTube on a regular basis and considers YouTube as the primary
online platform for video consumption\footnote{\url{https://www.youtube.com/watch?v=TUxlvb9Rcks&feature=youtu.be}}. The large
user base, broad geographic reach, and widespread adoption in the Indian subcontinent make YouTube a high quality source for the analysis in this paper.

Our data set was acquired using the following steps: (i) obtaining a set of search queries to execute against the YouTube search feature (ii) executing the searches against YouTube search to retrieve a list of relevant videos, (iii) crawling the comments for these videos using the publicly available YouTube API (further details presented in the Appendix).

\noindent \textbf{Collecting a set of queries:} We start with a seed set $\mathcal{S}$ of queries relevant to the crisis:
\texttt{[Pulwama]}, \texttt{[Balakot]}, \texttt{[Abhinandan]}, \texttt{[Kashmir]}, \texttt{[India Pakistan war], \texttt{[India Pakistan]}}\footnote{We noticed that the queries \texttt{[India Pakistan]} and \texttt{[Pakistan India]} yielded slightly different results, Following \cite{stephens2017everybody}, that revealed that we tend to put our more-preferred choice ahead in a pair, whenever we have a query that contained a country pair (e.g., \texttt{[India Pakistan]} or \texttt{[Pakistan India war]}, we adjusted the order of the pair accordingly matching it with the location of interest.}. We  construct \emph{News}, a set of highly popular news channels in India , Pakistan,
    and the world (listed in the Appendix). Next, we expand this query set and construct $\mathcal{S}_{related}$ by searching for each of the queries in $\mathcal{S}$ on Google Trends\footnote{\url{https://trends.google.com/trends/?geo=US}} setting the geographic location to India or Pakistan and including
    the related queries returned for the time duration of interest (14th February 2019 to 13th March 2019).  Finally, for each query $q$ in $\mathcal{S}_{related}$ and each channel $n$ in \emph{News}, a new query is formulated by
    concatenating $q$ and $n$. For instance, \texttt{[Pulwama CNN]} is obtained by concatenating the query \texttt{[Pulwama]}
    and news channel \texttt{[CNN]}. This final set of queries is called $\mathcal{S}_{final}$. Overall, $\mathcal{S}_{final}$ contains 6,210  queries of which 207  queries are obtained from Google Trends.

\noindent\textbf{Constructing $\mathcal{V}_{\emph{relevant}}$, a set of relevant videos:} For each query $q$ in $\mathcal{S}_{final}$, we execute a search using the YouTube search API to retrieving 200 most
relevant videos posted during the period of interest. This step yields a result set $\mathcal{V}$ (6,157 unique videos). after removing irrelevant (annotation criterion and details presented in extended version) and unpopular videos (less than 10 comments), we finally obtained $\mathcal{V}_{\emph{relevant}}$, a set of 2,890 videos. 

\noindent\textbf{Constructing $\mathcal{C}_{\emph{all}}$, the overall comments corpus:}
For each video $v$ in $\mathcal{V}_{relevant}$, the YouTube API is used to retrieve all the comments posted during the period of interest. The overall comments corpus, $\mathcal{C}_{\emph{all}}$, consists of 2,047,851 comments posted by 392,460 users.



\noindent\textbf{Constructing $\mathcal{C}_{\emph{en}}$, the English comments corpus:} We next extract English comments using a novel polyglot embedding based method first proposed in this paper (described in the results section). Our English comments corpus, $\mathcal{C}_{\emph{en}}$, consists of 921,235 English comments.  

\begin{figure}[t]
\centering
\fbox{\includegraphics[trim={5 5 5 5},clip, width=2.5in, height = 1.63in]{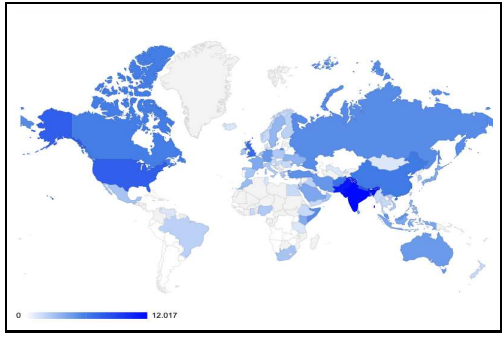}}
\vspace{0.5cm}
\caption{\small{Global participation log-scale choropleth. The darker regions indicate larger mention
counts (see, Table~\ref{tab:global-participation-count} for actual counts of top 10 countries). Countries in the Indian subcontinent and those with large Indian and Pakistani
populations feature heavily in the discussion.}}
\label{fig:globalPariticipation}
\end{figure}

\noindent\textbf{Investigating coverage:} It is important that the corpus reflects comments from both
conflicting countries. We conducted a text template-matching analysis to estimate the origin of the comments posted.
We manually inspected the corpus and observed that typically, origin and nationality were expressed through the following phrases: \texttt{[I'm], [I am], [I am from],[I am a],[I am an],[I am in],[I am in the],[I am from the]}, and \texttt{[love from]}. We used these templates  and retrieved five tokens following
each phrase. Country mentions are extracted from these following tokens and mention frequencies obtained. Using the above heuristics, overall, we assigned nationality to 7,806 users out of 392,460 users (1.99\%) accounting for 5.82\% of the comments present in the corpus. A log-scale choropleth visualization is shown in Figure~\ref{fig:globalPariticipation} and the 10 most mentioned countries are listed in Table~\ref{tab:global-participation-count}.

\begin{table}
\footnotesize
\centering
\begin{tabular}{|p{12cm}|}
\hline
\Tstrut
India (4,143), Pakistan (2,161), Bangladesh (345), Nepal (234), United states of America (163), United Kingdom (97), Afghanistan (85), China (66), Canada (57), Russia (35) \Bstrut\\ \hline
\end{tabular}
\caption{\small{Country with mention counts in brackets.}}
\label{tab:global-participation-count}
\end{table}

This analysis illustrates that our corpus contains (i) a balanced participation from both conflicting countries, and (ii) moderate participation from neighboring countries likely to be affected in the event of a war. Interestingly, the plot indicates
participation from nations with a significant Indian and Pakistani origin population (USA\footnote{\url{https://www.migrationpolicy.org/article/immigrants-us-states-fastest-growing-foreign-born-populations}}, United Kingdom\footnote{\url{http://worldpopulationreview.com/countries/united-kingdom-population/}}, South Africa). We conjecture that modern migration patterns, the nuclear arsenals of India and Pakistan,
and the broad global spread of Indian and Pakistani diaspora could be possible reasons for the global attention.

\section{Results and Analysis}

\subsection{Language identification}
Mining a multilingual corpus for insights requires separating out portions of the corpus written in distinct languages. This
is a critical step since annotators might be proficient in only a subset of the languages, and the majority of NLP tools are
designed for monolingual corpora. We now present an important result to navigate multilingual social media corpora like those
generated in the Indian subcontinent. 

\noindent\textbf{Polyglot word-embeddings discover language clusters.} Polyglot word-embeddings are real-valued word-embeddings obtained by training a single model on a multilingual corpus. Polyglot word-embeddings
have received attention recently for  demonstrating performance improvements across a variety of NLP tasks~\cite{mulcaire-etal-2019-polyglot, mulcaire-etal-2018-polyglot, mulcaire2019low}. 
While the downstream impact of the embeddings has been explored, in the context of language identification, we perform the first qualitative and quantitative
analysis of this embedding space for a variety of Indian and European languages and present the following observations:
(i) The word-embedding space is divided into highly-accurate language clusters, (ii) a simple algorithm like $k-$Means
can retrieve these clusters, and (iii) the quality of the resulting clusters is on-par with predictions made by large-scale supervised language-identification systems in some cases.

\begin{table}
  \tiny
  \centering
  \begin{tabular}{|l|c|l|c|c|c|}
    \hline
    Method & Accuracy & Language & P & R & F1 \TBstrut\\
    \hline
    Our method & \textbf{0.99} & Hindi (E) (52.5\%)  & \textbf{1.0} & \textbf{0.98}  & \textbf{0.99} \TBstrut\\
               &      & English  (46.5\%) & \textbf{0.99}  & \textbf{1.0}  & \textbf{0.99} \\
               &      & Hindi (1\%)  & \textbf{1.0} & \textbf{1.0}  & \textbf{1.0} \TBstrut\\
    \hline
    \texttt{fastTextLangID} & 0.48 & Hindi (E) (52.5\%)   & \textbf{1.0}   & 0.01  & 0.02 \\
             &       & English  (46.5\%) & 0.55  & \textbf{1.0}  & 0.71 \\
             &       & Hindi (1\%)  & \textbf{1.0}   & \textbf{1.0}   & \textbf{1.0} \TBstrut\\
    \hline
    \texttt{fastTextLangID}$_{\emph{fair}}$ & 0.48 & Hindi (E) (52.5\%)   & \textbf{1.0}   & 0.01  & 0.02 \\
             &       & English  (46.5\%) & 0.55  & \textbf{1.0}  & 0.71 \\
             &       & Hindi (1\%)  & \textbf{1.0}   & \textbf{1.0}   & \textbf{1.0} \TBstrut\\
    \hline
    \texttt{GoogleLangID} & 0.96 & Hindi (E) (52.5\%)   & 0.97 & 0.94 & 0.96 \\
                 &      & English  (46.5\%) & 0.97 & 0.97 & 0.97 \\
                 &      & Hindi (1\%)  & 0.4 & \textbf{1.0}  & 0.57 \TBstrut\\
    \hline
    \end{tabular}

   \caption{\small{Language written in Roman script is indicated with (E). 
    Percentage of the ground truth assigned this label is indicated for each language. Best metric is highlighted in bold for
    each language. P: precision, R: recall.}}
    \label{tab:kashmirResults}
\end{table}

\begin{figure}[t]
\centering
\includegraphics[trim={0 0 0 0},clip, width=3.1in, height = 2.3in]{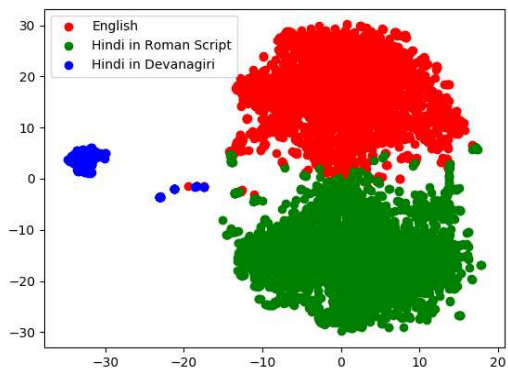}
\caption{\small{A visualization of the polyglot document-embedding space. The clusters are retrieved with $k$-means with $k$ set to 3 (see, Table~\ref{tab:kashmirResults} for empirical results). Additional experimental results covering a wider range of data sets and languages (European languages, low-resource Indian languages) are presented in the Appendix.}}
\label{fig:cluster}
\vspace{-0.3cm}
\end{figure}

For generating the embeddings, we first strip all punctuation and tokenize by splitting on whitespace.
Next, 100-dimensional FastText~\cite{bojanowski2017enriching} embeddings are trained on the full corpus yielding the polyglot embeddings. The FastText embeddings use the SkipGram training objective~\cite{DBLP:journals/corr/abs-1301-3781} where an input word's context is predicted. The model is parameterized by a set of real valued vectors (the word-embeddings) for each word in the vocabulary.
A full comment (document) embedding can be obtained by normalizing the word-embedding of each of the tokens in the comment and
subsequently averaging these word-embeddings. 

\noindent\textbf{Qualitative analysis:} A two-dimensional (2D) visualization of the document (comment) embedding space generated through applying the TSNE algorithm~\cite{maaten2008visualizing} on the computed document embeddings of a random sample of 10,000 comments is shown
in Figure~\ref{fig:cluster}. We observe three clusters in the visualization.
We then run $k$-Means on these document embeddings setting $k$ to 3 based on this observation. A manual inspection of the clusters reveals that they correspond to (i) Hindi in Roman script (green), (ii) Hindi in Devanagari script (blue), (iii) English (red).

\noindent\textbf{Quantitative analysis:} We next construct a technique for comment language identification. First, each
comment's embedding is obtained by the scheme described above. Next, the value of $k$ for the $k$-Means algorithm is chosen
using a standard heuristic~\cite{rousseeuw1987silhouettes} and $k$-Means is run which yields $k$ clusters. Finally, 
a sample of 10 comments is drawn from each of the obtained
clusters and the dominant language from this sample is assigned to this cluster. In our experience, at least 8 out of 10 comments in
the sample were from the dominant language i.e. each of the clusters obtained contains a highly dominant language
and the value of $k$ matches the number of languages present in the corpus.
A test comment is assigned a language by (i)
computing its embedding (as mentioned above), (ii) assigning this comment (embedding) to the cluster whose center is closest, (iii) returning the cluster's assigned language label.

We evaluate performance on a held-out set of 200 documents and report precision, recall, F1, and accuracy. 3 languages were discovered by annotators - English, Hindi in Roman script (denoted Hindi(E)), and Hindi written in Devanagiri (see, Table~\ref{tab:kashmirResults}). Note that, Hindi (mainly spoken in India) and Urdu (mainly spoken in Pakistan) are registers of the same language. Neither our annotators, nor commercial and open source solutions were able to distinguish between the two and thus the Hindi(E) cluster is used to denote both.  
We compare against two strong supervised baselines - (i) \texttt{fastTextLangID}\footnote{\url{https://fasttext.cc/docs/en/language-identification.html}} - a popular open source
solution supporting 174 languages, and (ii) \texttt{GoogleLangID}\footnote{\url{https://cloud.google.com/translate/docs/detecting-language}} - a commercial solution able to identify close to 100 languages. 

\noindent\textbf{Fairness:} We first emphasize that the main purpose of comparing against \texttt{fastTextLangID} and \texttt{GoogleLangID} is \textbf{not to claim our minimally supervised solution is superior to supervised solutions across the board} but \emph{to demonstrate our solution's effectiveness in the specific domain of noisy social media texts generated in the Indian subcontinent}. A fair performance comparison between the two supervised baselines and our proposed approach is challenging for several reasons. On one hand, due to varying levels of resources, supervised solutions might not be trained to predict the languages (expressed in non-native scripts) in the corpus. The baselines also predict from a larger set of languages. In contrast, our method reveals only those languages observed in the corpus in question - thus a limited set of clusters (labels) is obtained - in most cases this is substantially smaller than the number of languages supported by industrial strength baselines. On the other hand, the baselines are supervised methods that have been trained on vast amounts of annotated data whereas our methods require minimal manual labeling - a critical feature for dealing
with corpora featuring low resource languages which are a common occurrence in the Indian subcontinent.

We agree that restricting the baselines to predict only from the smaller set may offset the advantage of our method. The API for \texttt{fastTextLangID} provides an ordered list of all languages that it supports with the confidence score (\texttt{GoogleLangID} does not provide this feature). Let the set of all languages present in a corpus be denoted as $\mathcal{L}$. For a given document, we predict the language belonging to $\mathcal{L}$ with the highest confidence score. Suppose the top three predictions for a document from our data set by \texttt{fastTextLangID} are (1) \emph{German} (predicted with highest confidence), (2) \emph{Spanish} and (3) \emph{Hindi}. Since \emph{Hindi} $\in \mathcal{L}$, and \emph{German} $\notin \mathcal{L}$, \emph{Spanish} $\notin \mathcal{L}$, we consider that the predicted label is \emph{Hindi}. We denote this new setting as \texttt{fastTextLangID}$_{\emph{fair}}$.

Results are present in Table~\ref{tab:kashmirResults}. We observe that our method and the \texttt{GoogleLangID} are able to achieve near-perfect
results while both \texttt{fastTextLangID} and \texttt{fastTextLangID}$_{\emph{fair}}$ mislabel the Hindi(E) cluster comments as English underscoring the importance of our method. We re-iterate that the purpose of this
analysis is to illustrate that when low-resource multilingual settings are encountered, large-scale supervised solutions
might not be capable of supporting the desired analyses; our minimal supervision
method produces outcomes with reasonable accuracy and avoids imposing significant annotation burdens.

A thorough treatment of this data set, and additional data sets containing a variety of Indian languages (low-resource languages Bengali and Oriya)  and European languages (21 languages) is presented in the Appendix.

\noindent\textbf{Intuition:} The SkipGram model used for training the FastText embeddings predicts a word's context given a word. In a polyglot setting, the likeliest context predicted for a Hindi word is other Hindi
words. The embeddings likely reflect this aspect of the language model and
thus we see language clusters. We admit that implementation choices like
splitting on whitespace (for instance) can preclude some languages so
we refrain from making claims about the universality of the technique and
present empirical results only on Indian and European languages.

\subsection{Temporal trends in pro-peace intent} 
State of the art sentiment analysis tools typically target domains like movie reviews, product reviews and so on. Prior sentiment analysis research has been been performed on political news content~\cite{kaya2012sentiment} and social media responses to humanitarian crises~\cite{ozturk2018sentiment}, but to the best of our knowledge, there has been no previous work on war sentiment. Moreover, most of these standard off-the-shelf sentiment analysis tools have been trained on corpora very different from ours. For instance, OpenAI sentiment analysis tool~\cite{radford2017learning} is trained on Amazon e-commerce product reviews. Consequently, off-the-shelf tools are not sufficient in our case. For instance, Stanford CoreNLP (version 3.9.2) sentiment analysis\footnote{\url{https://corenlp.run/}}~\cite{manning-EtAl:2014:P14-5}, a popular sentiment analysis model, marks the following three examples: \texttt{[Say no to war.]}, \texttt{[War is not a solution.]}, and \texttt{[We will nuke you.]} as negative, neutral, and positive, respectively. In a conflict-analysis scenario, these three examples should be marked as positive, positive, and negative instead. Moreover, we observed that the predicted results are sensitive to punctuation and casing - which cannot be guaranteed in a noisy setting. Hence, we address the challenges in modeling sentiment in our corpus by using a comprehensive manually labeled set of phrases to reveal sentiment. Techniques for analyzing the semantic orientation of text have heavily exploited manually curated lexicons~\cite{taboada2011lexicon,velikovich2010viability,hamilton2016inducing,o2010tweets}. Following~\cite{velikovich2010viability,hamilton2016inducing}, we construct an annotated domain-specific phrase lexicon for mining pro-war and pro-peace intent.

\begin{figure*}
\centering
\subfigure[Comment activity over time]{%
\includegraphics[trim={0 0 0 135},clip, width = 0.45 \textwidth]{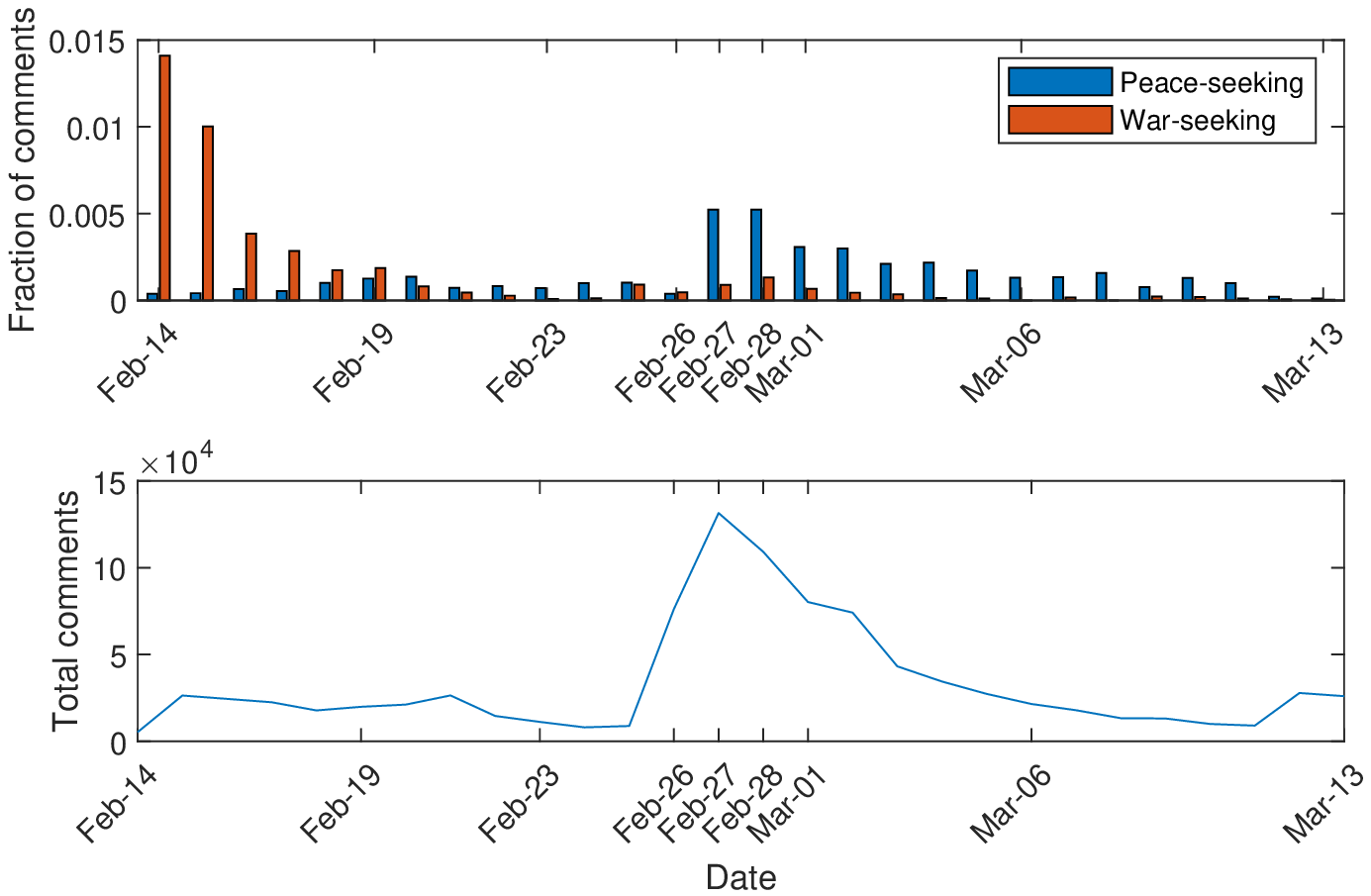}
\label{fig:commentActivity}}
\subfigure[Like activity over time]{%
\includegraphics[trim={0 0 0 135},clip, width = 0.45 \textwidth]{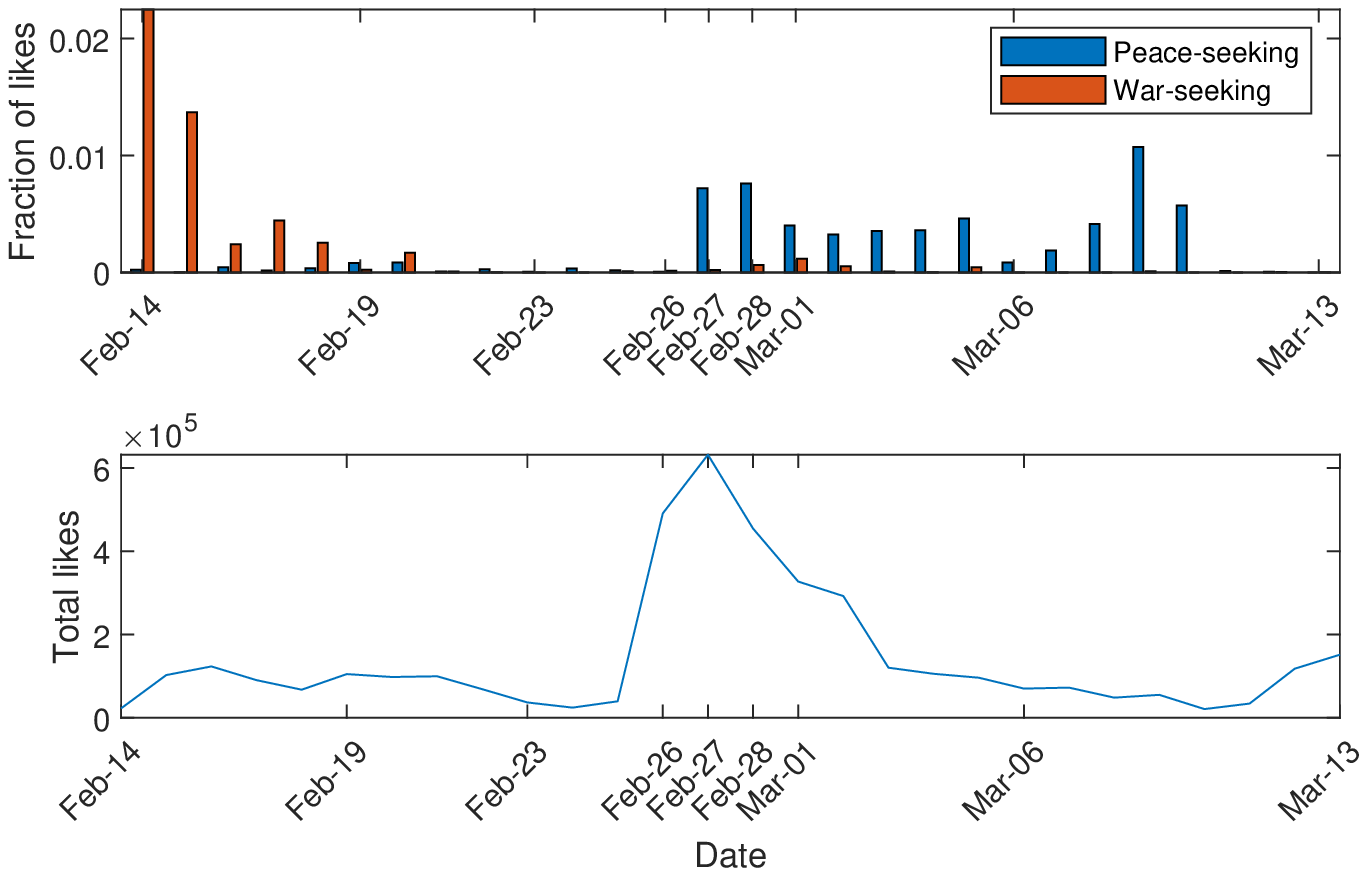}
\label{fig:LikeActivity}}
\subfigure[Comments with commonly used intent phrases]{%
\includegraphics[trim={0 135 0 0},clip, width = 0.45 \textwidth]{CommentFourPhrases.eps}
\label{fig:fourPhraseComment}}
\subfigure[Likes for such comments]{%
\includegraphics[trim={0 135 0 0},clip, width = 0.45 \textwidth]{LikeFourPhrases.eps}
\label{fig:fourPhraseLike}}
\subfigure[Comments with comprehensive intent phrases]{%
\includegraphics[trim={0 135 0 0},clip, width = 0.45 \textwidth]{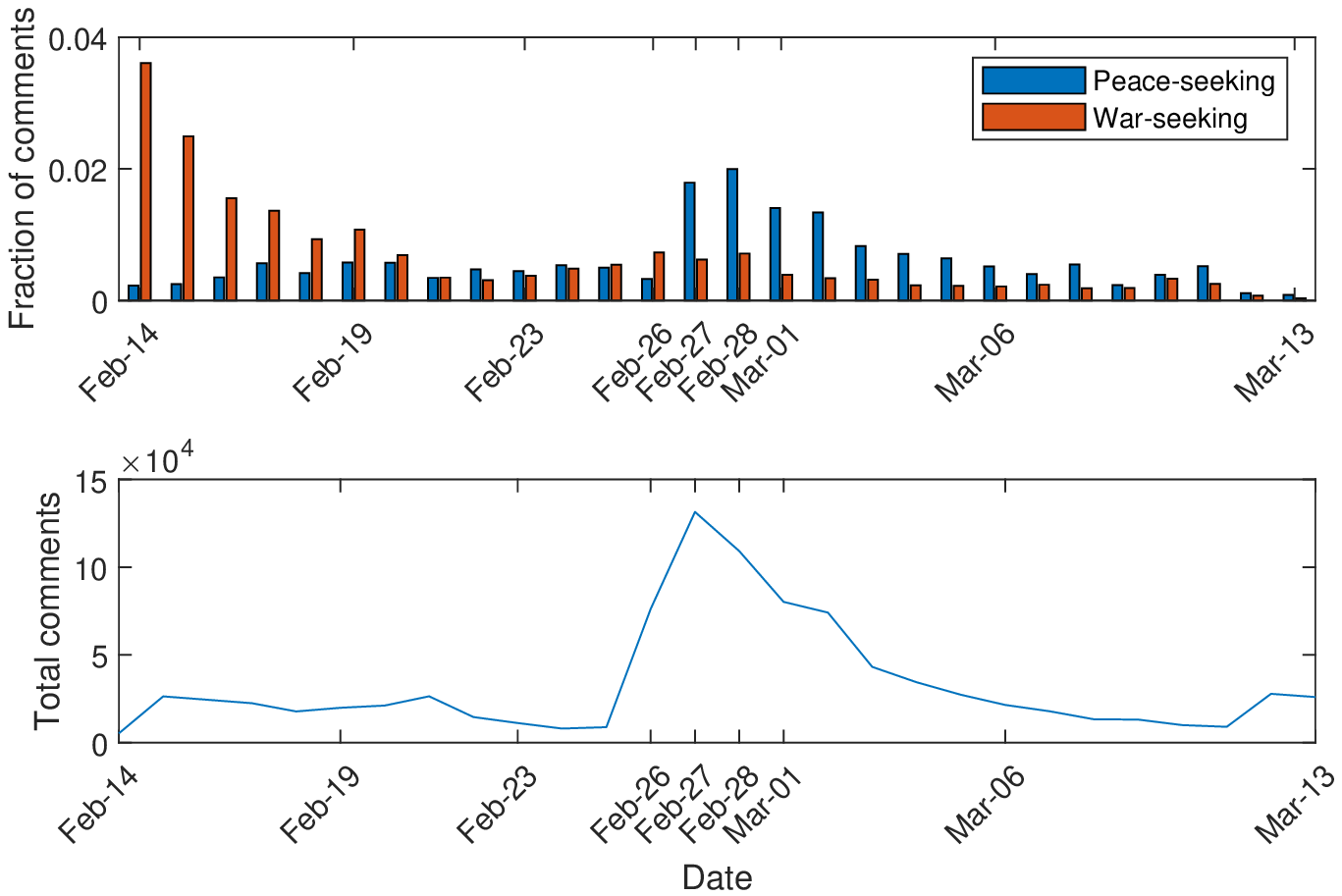}
\label{fig:allPhraseComment}}
\subfigure[Likes for such comments]{%
\includegraphics[trim={0 135 0 0},clip, width = 0.45 \textwidth]{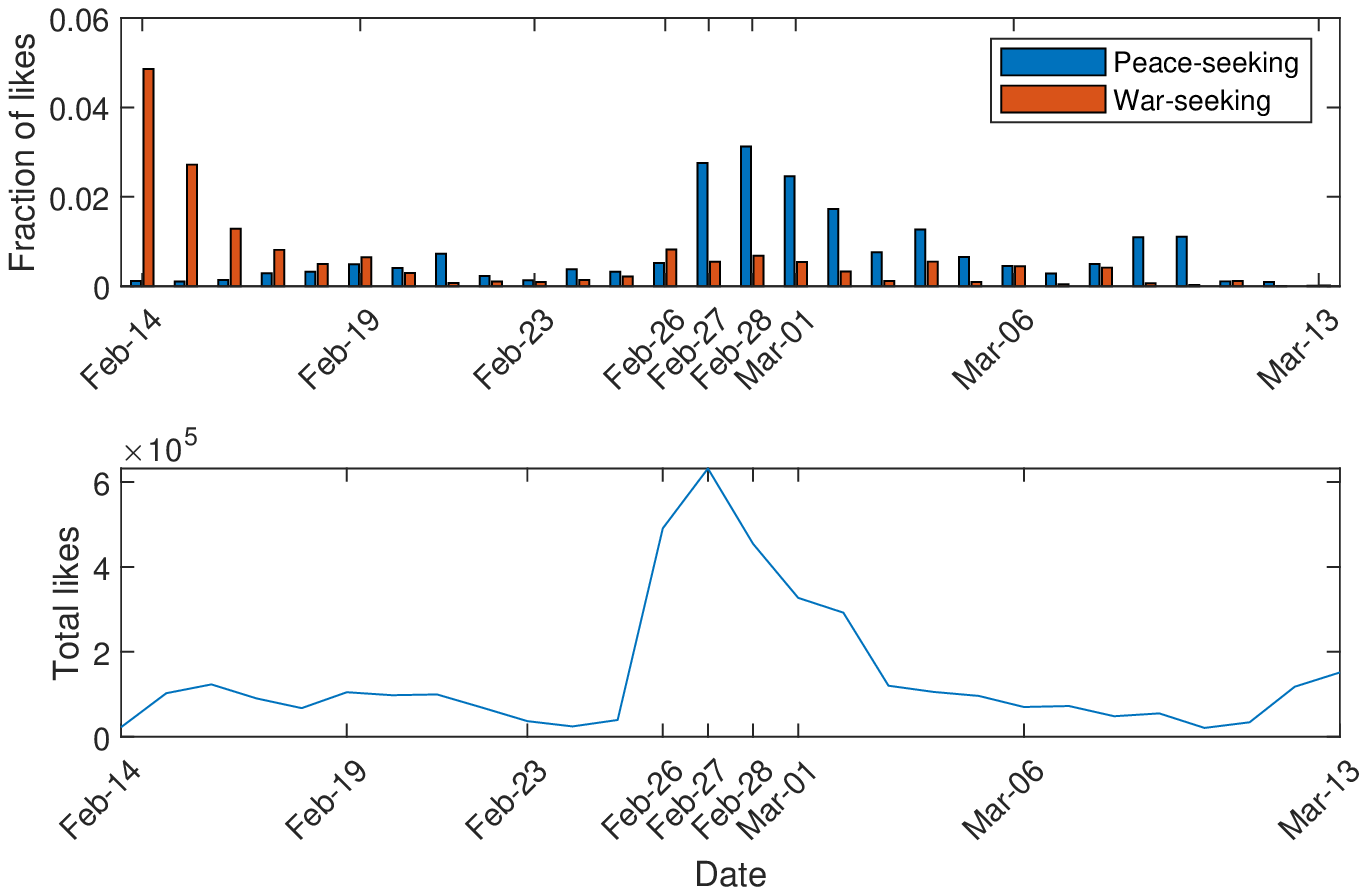}
\label{fig:allPhraseLike}}
\caption{\footnotesize{Temporal shift of pro-war and pro-peace intent.}}
\label{fig:intentShift}
\end{figure*}


We first identify a set of four high-frequency trigrams expressing collective war/peace intent: [\texttt{we want peace}], [\texttt{we want war}], [\texttt{we want surgical}] (\emph{surgical} refers to surgical strike), [\texttt{we want revenge}]. These express peace, war, war, and war intents respectively. Out of 9,300,740 unique trigrams, these four trigrams are the 35$^{th}$, 515$^{th}$, 875$^{th}$ and 967$^{th}$  in terms of frequency and are the top four collective intent expressing trigrams (trigrams that start with “we” followed by a volitional verb; e.g., [\texttt{we want}], [\texttt{we need}]). A comment that contains $m$ instances of a peace-seeking (war-seeking) phrase receives a positive (negative) score of $m$ ($n$). The overall score of a comment is $m - n$. The comment expresses peace-seeking intent if the overall score is greater than 0, neutral intent if the overall score is equal to 0 and war-seeking intent if the overall score is less than 0. 

We summarize the temporal trends of peace-seeking and war-seeking intent using the four frequently used trigrams in Figure~\ref{fig:fourPhraseComment} and \ref{fig:fourPhraseLike}. We normalize war and peace intent frequencies by the total number of likes or comments received on that day, giving us values in the [0,1] interval, and allowing us to compare activities and sentiment across different days. We measure engagement in terms of comments and likes and plot the overall comment activity (Figure~\ref{fig:fourPhraseComment}) and overall like activity (Figure~\ref{fig:fourPhraseLike}) along with the respective temporal trend plots. As shown in Figure~\ref{fig:fourPhraseComment} and \ref{fig:fourPhraseLike}, the baseline user activity, both in terms of comments and likes, spiked around the \texttt{IAFPILOT-CAPTURE} and \texttt{IAFPILOT-RELEASE} events (nearly 6 times more user engagement on 27$^{th}$ as compared to 15th$^{th}$). 
Figure~\ref{fig:fourPhraseComment} and Figure~\ref{fig:fourPhraseLike} show that right after \texttt{PULWAMA}, pro-war intent dominated pro-peace intent. Following the pilot's capture and subsequent release declaration, there was a substantial shift towards pro-peace intent after which, the pro-peace intent generally dominated war-seeking intent. Feb 27$^{th}$ was also the day when Pakistan media reported a meeting between the Pakistan PM and the nuclear warheads body and several news videos discussed the possibility of a nuclear war. Human evaluation on randomly sampled 200 positive comments reveals the following takeaways: in the context of this particular conflict, (1) pro-peace intent spiked when the possibility of a war became real, and (2) the peace-gesture by the Pakistan Government possibly influenced this shift as Indians deeply cared about their captured pilot's safety and appreciated the pilot's safe return.

In order to widen our coverage, we constructed an extensive lexicon of polarity phrases (sample phrases are listed in Table~\ref{tab:samplePhrase} with in-depth description presented in~\cite{kashmir}). Overall, we obtained 3,104 annotated phrases as one of: (i) peace-seeking (310 phrases), (ii) war-seeking (278 phrases), or (iii) neutral or unclear (2,516 phrases). Our annotators were instructed to label explicit calls for war and peace. Similar to our previous setting, for a given comment, presence of a peace-seeking phrase contributes +1 to the comment's score, a war-seeking phrase contributes -1 to the overall score, an a neutral phrase contributes a score of 0.
The longest matching phrase is considered for computing the sentiment score and all subsumed phrases are ignored. For instance, consider a comment \texttt{[we want peace but India is not worth it]}; if [\texttt{we want peace}] has score +1 and [\texttt{we want peace but India}] has score 0, [\texttt{we want peace}] is disregarded and the overall contribution from these 2 phrases is 0.

\begin{table}
{
\begin{center}
     \begin{tabular}{| l  | l |}
     \hline 
     Pro-peace & Pro-war \\
    \hline
     
   war is not a solution & we are ready for war \\
   \hline
   war is not the solution & war is the only solution \\
   \hline
   we want peace not war & we are ready to fight \\
   \hline
   we dont want war we  & we are ready to die \\
   \hline
    say no to war & nuke pakistan \\
   \hline
  peace between india and pakistan & nuke the shit out of \\
    \hline
     peace between pakistan and india  & want to go to war \\
   \hline
   want peace in both countries  & war is the only option \\
   \hline
    pakistan wants peace with india & wipe out pakistan from the \\
   \hline
   war is not a joke & wipe india off the map \\
   \hline
\end{tabular}

\end{center}
\caption{{A random sample of 10 pro-peace and 10 pro-war phrases.}}
\label{tab:samplePhrase}}
\end{table}

As shown in Figure~\ref{fig:allPhraseComment} and Figure~\ref{fig:allPhraseLike}, the qualitative trends found in our previous analysis hold. Right after \texttt{PULWAMA}, pro-war intent dominated  pro-peace intent and a visible shift was observed on and after Feb 27th. Additionally, our coverage (fraction of comments containing at least one intent-expressing phrase) improved; overall, we obtained 7.25\% coverage of comments (20x more than before) and 10.42\% (24x more than before) coverage of likes. 

\noindent\textbf{\emph{Did many people change their minds?}} Unlike YouTube comments, YouTube likes are anonymous and cannot be attributed to individual users. 
Hence, we focus on the following research question: \emph{where there many users who initially clamored for war but later changed their minds? Or, when war became an imminent possibility, did a different sub-population voice their concerns?} Analysis reveals the latter case to be true. On our comprehensive intent-expressing phrase set, we found that 4,407 users posted one or more peace-seeking comments, while 7,402 users posted one or more war-seeking comments. 280 users posted both types of comments. The Jaccard index\footnote{defined as $\frac{|A \cap B|}{|A \cup B|}$ for sets $A$ and $B$} between the two user sets was 0.02 indicating low overlap. 




\noindent \textbf{Focused analysis around the peace-spike:} We now focus on a comparative analysis between the two time intervals when war (or revenge) and peace intents were at their respective maximums: a three day period starting on \texttt{PULWAMA} (denoted as \emph{war-spike}), and a three day period starting on \texttt{IAFPILOT-CAPTURE} (denoted as \emph{peace-spike}). We compute the respective unigram distributions $\mathcal{P}_{\emph{war-spike}}$ and $\mathcal{P}_{\emph{peace-spike}}$. Next, for each token $t$, we compute the scores $\mathcal{P}_{\emph{war-spike}}(t) - \mathcal{P}_{\emph{peace-spike}}(t)$, and $\mathcal{P}_{\emph{peace-spike}}(t) - \mathcal{P}_{\emph{war-spike}}(t)$ and obtain the top tokens ranked by these scores (indicating increased usage in the respective periods of interest). As listed in Table~\ref{tab:movers}, both \texttt{war} and \texttt{peace} were heavily used tokens during the \emph{peace-spike}. However, \texttt{war} was predominantly used in the context of peace (e.g., \texttt{[war is not a solution]}, \texttt{[we don't want war]}). Several users also identified themselves as Indian or Pakistani and expressed love for the neighbor country. During the \emph{war-spike}, demands for revenge, or a surgical strike, or an attack on Pakistan dominated. Heavy use of Kashmir specific keywords during the \emph{war-spike} and greater emphasis at the country level at the later stage was also consistent with the sequence of events that started as a regional terror attack and snowballed into an international crisis between two nuclear adversaries. We conducted a similar analysis on the set of Hindi comments and our observations align with English corpus. 

\begin{table}

{

\begin{center}
     \begin{tabular}{| p{6cm}  | p{6cm} |}
    \hline
    \Tstrut More presence during \emph{war-spike}  & More presence during \emph{peace-spike} \\
     \hline \Tstrut
 Islam,
Kashmiris, 
Muslim,
need,
people,
religion,
Muslims,
sad, 
Modi, 
China, 
\textcolor{red}{strike}, 
kill, 
soldiers, 
Kashmiri, 
rip, 
\textcolor{red}{revenge}, 
time, 
\textcolor{red}{surgical}, 
\textcolor{red}{attack}, 
Kashmir \Bstrut&      Pakistan, 
pilot, 
\textcolor{blue}{war}, 
media, 
India, 
pak, 
\textcolor{blue}{peace}, 
Imran, 
\textcolor{blue}{Indian}, 
\textcolor{blue}{Pakistani}, 
\textcolor{blue}{love}, 
fake, 
khan, 
shot, 
Abhinandan, 
air, 
f16, 
sir, 
video, 
mig \\
    \hline
    \end{tabular}
\end{center}
\caption{\footnotesize{Biggest shift in token usage in the three day period starting from \emph{war-spike}  and \emph{peace-spike}}.}
\label{tab:movers}}
\end{table}

\begin{table}[htb]
{
\begin{center}
     \begin{tabular}{|l | c | c | c | c |}
    \hline
    Features & Precision & Recall & F1 & AUC  \\
    \hline
   n grams & 81.95 $\pm$ 2.61\% &  74.61 $\pm$ 3.02\% & 78.07 $\pm$ 2.27\% & 94.62 $\pm$ 0.81\\
    \hline
   n grams + I & 81.99 $\pm$ 2.58\%  & 73.64 $\pm$ 2.87\% & 77.56 $\pm$ 2.19\% & 94.45 $\pm$ 0.84\\
    \hline
n grams + I + FT  & \textbf{82.01 $\pm$ 2.59\%}  & \textbf{75.36 $\pm$ 3.01\%} &  \textbf{78.51 $\pm$ 2.24\%} & \textbf{95.48 $\pm$ 0.68}\\
    \hline
    
    \end{tabular}

\end{center}
\caption{\emph{Hope-speech} classifier performance.}
\label{tab:classifier}}
\vspace{-0.4cm}
\end{table}

\subsection{\emph{Hope-speech} detection}~\label{sec:hopeSpeech} Analyzing and detecting hate-speech and hostility in social media~\cite{del2017hate, davidson2017automated, chandrasekharan2017you, dinakar2012common, liu2018forecasting} have received considerable attention from the research community. Hate-speech detection and subsequent intervention (in the form of moderation or flagging a user) are crucial in maintaining a convivial web environment. However, in our case where the civilians of two conflicting nations are engaging in heated discussions in a politically tense situation, detecting comments that can potentially diffuse hostility and bring the two countries together has particular importance, for instance by highlighting such comments or otherwise giving them more prominence.

 \noindent\textbf{Definition 1:} A comment is marked as  \emph{hope-speech}, if it exhibits any of the following:
\begin{compactenum}
\item  The comment explicitly mentions that the author comes from a neutral country (e.g., [\texttt{great job thanks from Bangladesh make love not war}]), and exhibits a positive sentiment towards both countries in the conflict.
\item The comment explicitly mentions that the author comes from one of the conflicting countries, and exhibits a positive sentiment to an entity (all people, media, army, government, specific professionals) of the other country (e.g., [\texttt{I am from Pakistan I love India and Indian people}]).

\item The comment explicitly urges fellow citizens to de-escalate, to stay calm.
\item The comment explicitly mentions that the author comes from one of the conflicting countries, and criticizes some aspect of the author's own country (e.g.,[\texttt{I am from India but Indian media very very bad}]). 
\item The comment criticizes some aspect of both of the conflicting countries. 
\item The comment urges both countries to be peaceful.
\item The comment talks about the humanitarian cost of war and seeks to avoid civilian casualties (e.g., [\texttt{peace is better than war as the price of war is death of innocent peoples}]).
\item The comment expresses unconditional peace-seeking intent (e.g., [\texttt{we want peace}]).
\end{compactenum}

\noindent If any of the following criteria are met, the comment is \textbf{\emph{not}} \emph{hope-speech}:

\begin{compactenum}
\item The comment explicitly mentions that the author comes from a conflicting country and expresses no positive sentiment toward the other conflicting country. 
\item The comment explicitly mentions that the author comes from a neutral country but takes a position favoring only one of the conflicting countries (e.g., [\texttt{I m frm Australia I support Pakistan}]).
\item The comment actively seeks violence (e.g., [\texttt{I want to see Hiroshima and Nagasaki type of attack on Pakistan please please}]). 
\item The comment uses racially, ethnically or nationally motivated slurs (e.g., porkistan, randia).
\item The comment starts the proverbial whataboutism, i.e., we did \emph{b} because you did \emph{a} (e.g., [\texttt{Pakistan started it by causing Pulwama attack killing 44 Indian soldiers}]). 
\end{compactenum}

Our list of hostility diffusing criteria is not exhaustive and may not cover the full spectrum of hostility diffusing comments. Consequently, we agree that it is possible to have several other reasonable formulations of \emph{hope-speech}. Also, in a conflict scenario involving more than two conflicting entities, this particular definition may not hold. However, upon manual inspection of the corpus, we found that the definition covers a wide range of potentially hostility-diffusing comments while capturing several nuances.

\noindent\textbf{Hope-speech comment frequency in the wild:} On 2000 randomly sampled comments (500 from each week), our annotators found 49 positives (2.45\%), 1946 negatives and 5 indeterminate comments. This indicates that detecting \emph{hope-speech} is essentially a rare positive mining task which underscores  automated detection's importance.

\noindent\textbf{Training set construction using Active Learning:} To ensure generalizability and performance in the wild, it is critical that the training set contains sufficient examples from both classes and captures a wide variety of data points. To ensure this, we divided the corpus into four weekly sub-corpora and sampled uniformly from each of these acknowledging the strong temporal aspect in our data; for a data set consisting of sufficient number of positives and negatives, we employed a combination of Active Learning strategies~\cite{sindhwani2009uncertainty} and constructed a data set of 2,277 positives and 7,716 negatives (detailed description is presented in the Appendix). All rounds of manual labeling were performed by two annotators proficient in English. The annotators were presented with the definition and a small set of examples. They were first asked to label independently, and then allowed to discuss and resolve the disagreed labels. We obtained strong agreement in every round (lowest Cohen's $\kappa$ coefficient across all rounds was 0.82 indicating strong inter-rater agreement).

\noindent\textbf{Features:} We considered the following features: 
\begin{compactenum}
\item n-grams up to size 3 following existing literature on text classification~\cite{manning1999foundations}. 
\item the previously described sentiment score of a comment obtained using our comprehensive set of intent phrases (denoted as I in Table~\ref{tab:classifier}).
\item the previously described 100-dimensional polyglot FastText embeddings (denoted as FT in Table~\ref{tab:classifier}).
\end{compactenum}

\begin{figure}[t]
\centering
\includegraphics[trim={0 0 0 0},clip, width=2.9in, height = 1.8in]{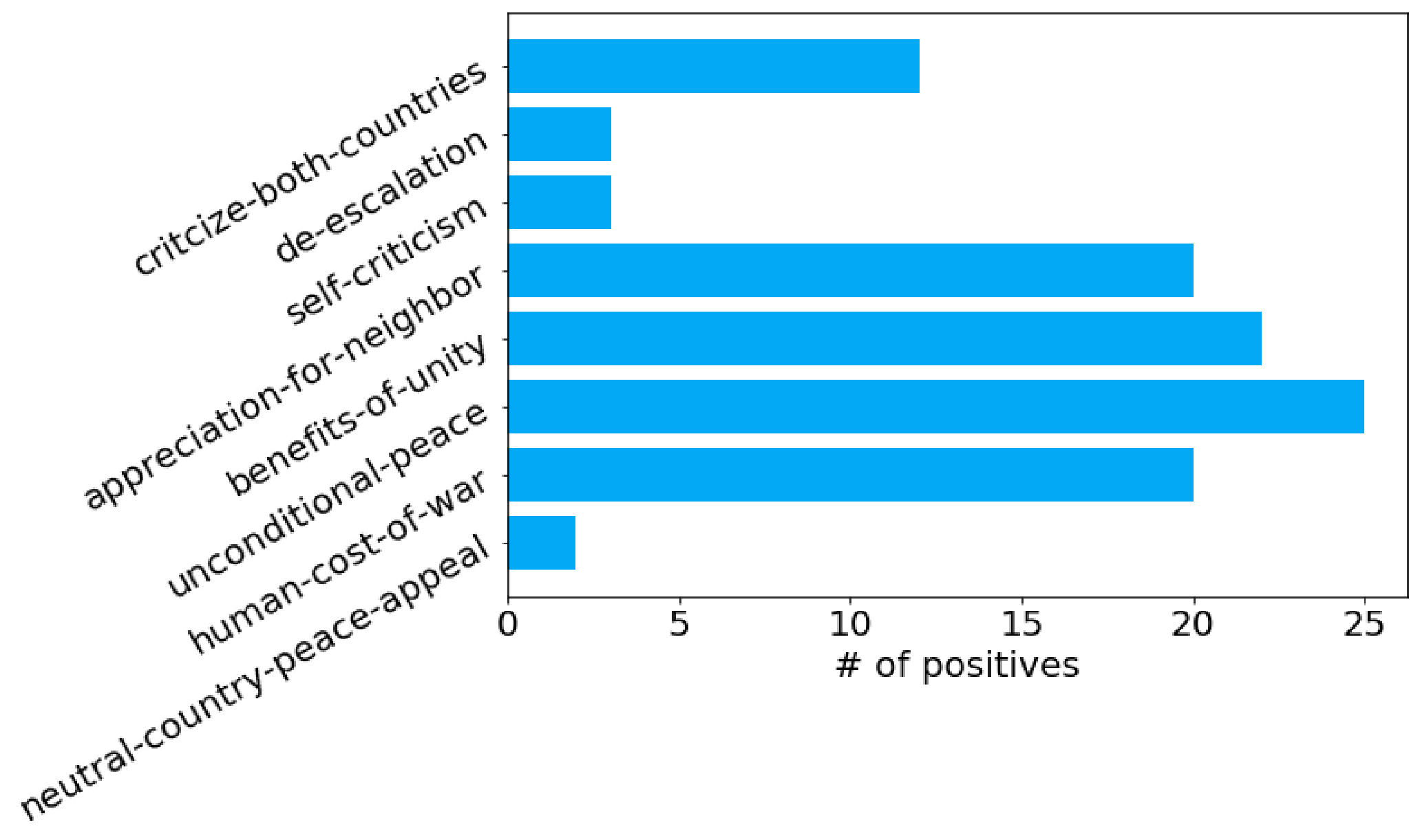}
\caption{\footnotesize{Breakdown of positive comments found in the wild. A single comment can satisfy multiple criteria.}}
\label{fig:distribution}
\vspace{-0.2cm}
\end{figure}

\noindent\textbf{Classifier performance:} On our final data set, we used a 80/10/10 train/validation/test split. On the training set, we train a logistic regression classifier with L2 regularization with the discussed features and report performance on the test set. The experiment was run 100 times on 100 randomly chosen splits. As shown in Table~\ref{tab:classifier}, the results indicate that a \emph{hope-speech} classifier with good precision and recall can be constructed. We admit that off-the-shelf sentiment analysis tools may perform poorly in our task, and it is not a fair comparison since they are trained for a different domain. However, for the sake of completeness, we ran the Stanford CoreNLP sentiment analyzer on our data set (precision: 27.65\%, recall: 41.45\%, F1: 33.17\%). Our baselines' stronger performance  indicates that the task of \emph{hope-speech} detection is different from simple sentiment analysis and hence requires a targeted approach.  

\begin{table}
{
\scriptsize
\begin{center}
     \begin{tabular}{|p{13cm}|}
    \hline 
    +Abbas Ali look bro you don't realize a girl's agony I have lost my brother . And today is my birthday. I have lost the peace of my life in the rage of war. I have left with no one to atleast make me happy. please I request you are just like my brother stop talking about  war for the humanity sake please bhaiya stop this\\
    \hline 
    I am Iranian, I don't like two nuclear powers go to war with each other right next to us! Please make peace or all of us will suffer.\\
    \hline
    Say no to any war please. We need world without war. \\
    \hline
 India and Pakistan should stop fighting. Till when we will keep fighting?? Humanity should win.  Let's find peace and move towards development.  \\
    \hline
 Let the indians and Pakistanis come together and spread peace. Nobody wants war. We all know what happened in Afghanistan, in Iraq, in Syria. What wounds inflicted by ISIS on the followers of Islam and other religions. Terrorists have no religion. Any religion never spreads hatred. Politicians are spreading hatred. So they also do not have any religion.  We are educated mass. We were a single country.  No offence for the mistaken identity of the lady. Let's forget our religion and work for greater peace. Hindustan zindabad and Pakistan zindabad. Terrorism murdabad   \\
   \hline 
   I am from Pakistan. We don't need any kind of support from anywhere. India is our neighbor and we respect Indians and humanity all around the world. Please don't spread haters both of us don't want war. We are neighbors and love each other. Long live Pakistan $\&$ India.\\
    \hline
    Comments are filled with love. Both the countries do not want any war, and both of us are seeking peace. In spite of that why is this happening? Why the attacts are taking place? Why is there is a heated situation between both the countries. Unfortunately politics is still successful in maintaining the division among Hindustan-Pakistan (divide $\&$ rule)\\
    \hline
   Thank you Pakistani soldiers for good care of our soldier send him back safely to india\\
   \hline 
   as a pakistani i agree and i think we indo pak peoples condemn this  war .\\
    \hline
    Im admiring u sir wen u wer in cricket, u wer the best captain in Pakistan team till now, after dat no one came like u, same way plz try to stop all this war talk ASAP, we love peace as much as u do, sit and talk and dissolve this.... Love n peace from India.\\
    \hline 
    
    \end{tabular}
\end{center}
\caption{\footnotesize{Random sample of comments in the wild marked as \emph{hope-speech} by our classifier.}}
\label{tab:samplesWild}}
\end{table}

\noindent\textbf{Performance in the wild:} We randomly sampled 1000 unlabeled comments from each day and ran our \emph{hope-speech} classifier. Overall, 111 comments were predicted as positives with 94 verified correct by human evaluation (precision: 84.68\%). Recall that, simple random sampling uncovered 2.45\% of comments exhibiting \emph{hope-speech}. Hence, our performance in the wild holds promise for substantially reducing manual moderation effort (example comments are presented in Table~\ref{tab:samplesWild}). We further analyzed the sub-categories of positives found in the wild. As shown in Figure~\ref{fig:distribution}, our \emph{hope-speech} classifier was able to find all different sub-categories of positive comments.

\section{Conclusion}

In the era of ubiquitous internet, public opinion on a rapidly evolving global issue can exhibit similar fast-changing behavior, much of which is visible to a very large fraction of internet users. Consequently, this poses an additional challenge to countries with a history of past conflicts as comments inciting hostility may spiral the public opinion towards a stronger pro-war stance. In this work, we define a novel task of \emph{hope-speech} detection to identify hostility-diffusing content.  Extreme web-moderation during periods of strife and tension has included 
completely disabling internet access in a locality. Our work in detecting hostility-diffusing content may find applications in these scenarios as well. We present a thorough analysis of a novel polyglot embedding based language identification module that can be useful in facilitating research on social media data generated in this part of the globe with presence of several low-resource languages.

\bibliographystyle{unsrt}


\begin{thebibliography}{10}

\bibitem{schofield2010kashmir}
Victoria Schofield.
\newblock {\em Kashmir in conflict: India, Pakistan and the unending war}.
\newblock Bloomsbury Publishing, 2010.

\bibitem{toon2019rapidly}
Owen~B Toon, Charles~G Bardeen, Alan Robock, Lili Xia, Hans Kristensen, Matthew
  McKinzie, RJ~Peterson, Cheryl~S Harrison, Nicole~S Lovenduski, and Richard~P
  Turco.
\newblock Rapidly expanding nuclear arsenals in pakistan and india portend
  regional and global catastrophe.
\newblock {\em Science Advances}, 5(10):eaay5478, 2019.

\bibitem{zeitzoff2017social}
Thomas Zeitzoff.
\newblock How social media is changing conflict.
\newblock {\em Journal of Conflict Resolution}, 61(9):1970--1991, 2017.

\bibitem{gochman1983realpolitik}
Charles~S Gochman and Russell~J Leng.
\newblock Realpolitik and the road to war: An analysis of attributes and
  behavior.
\newblock {\em International Studies Quarterly}, 27(1):97--120, 1983.

\bibitem{malik2002kashmir}
Iffat Malik and Robert~G Wirsing.
\newblock {\em Kashmir: Ethnic conflict international dispute}.
\newblock Oxford University Press Oxford, 2002.

\bibitem{bose2009kashmir}
Sumantra Bose.
\newblock {\em Kashmir: Roots of conflict, paths to peace}.
\newblock Harvard University Press, 2009.

\bibitem{staniland2013kashmir}
Paul Staniland.
\newblock Kashmir since 2003: Counterinsurgency and the paradox of
  “normalcy”.
\newblock {\em Asian Survey}, 53(5):931--957, 2013.

\bibitem{ali1983can}
Tariq Ali.
\newblock {\em Can Pakistan survive?: the death of a state}.
\newblock Penguin Books London, 1983.

\bibitem{pandya2019future}
Abhinav Pandya.
\newblock The future of indo-pak relations after the pulwama attack.
\newblock {\em Perspectives on Terrorism}, 13(2):65--68, 2019.

\bibitem{feyyaz2019contextualizing}
Muhammad Feyyaz.
\newblock Contextualizing the pulwama attack in kashmir--a perspective from
  pakistan.
\newblock {\em Perspectives on Terrorism}, 13(2):69--74, 2019.

\bibitem{mulcaire-etal-2018-polyglot}
Phoebe Mulcaire, Swabha Swayamdipta, and Noah~A. Smith.
\newblock Polyglot semantic role labeling.
\newblock In {\em Proceedings of the 56th Annual Meeting of the Association for
  Computational Linguistics (Volume 2: Short Papers)}, pages 667--672,
  Melbourne, Australia, July 2018. Association for Computational Linguistics.

\bibitem{mulcaire-etal-2019-polyglot}
Phoebe Mulcaire, Jungo Kasai, and Noah~A. Smith.
\newblock Polyglot contextual representations improve crosslingual transfer.
\newblock In {\em Proceedings of the 2019 Conference of the North {A}merican
  Chapter of the Association for Computational Linguistics: Human Language
  Technologies, Volume 1}, pages 3912--3918, June 2019.

\bibitem{mulcaire2019low}
Phoebe Mulcaire, Jungo Kasai, and Noah~A Smith.
\newblock Low-resource parsing with crosslingual contextualized
  representations.
\newblock In {\em Proceedings of the 23rd Conference on Computational Natural
  Language Learning (CoNLL)}, pages 304--315, 2019.

\bibitem{schmidt2017survey}
Anna Schmidt and Michael Wiegand.
\newblock A survey on hate speech detection using natural language processing.
\newblock In {\em Proceedings of the Fifth International Workshop on Natural
  Language Processing for Social Media}, pages 1--10, 2017.

\bibitem{hessel-lee-2019-somethings}
Jack Hessel and Lillian Lee.
\newblock Something{'}s brewing! early prediction of controversy-causing posts
  from discussion features.
\newblock In {\em Proceedings of the 2019 Conference of the North {A}merican
  Chapter of the Association for Computational Linguistics: Human Language
  Technologies, Volume 1 (Long and Short Papers)}, pages 1648--1659,
  Minneapolis, Minnesota, June 2019. Association for Computational Linguistics.

\bibitem{stephens2017everybody}
Seth Stephens-Davidowitz and Andr{\'e}s Pabon.
\newblock {\em Everybody lies: Big data, new data, and what the internet can
  tell us about who we really are}.
\newblock HarperCollins New York, 2017.

\bibitem{bojanowski2017enriching}
Piotr Bojanowski, Edouard Grave, Armand Joulin, and Tomas Mikolov.
\newblock Enriching word vectors with subword information.
\newblock {\em Transactions of the Association for Computational Linguistics},
  5:135--146, 2017.

\bibitem{DBLP:journals/corr/abs-1301-3781}
Tomas Mikolov, Kai Chen, Greg Corrado, and Jeffrey Dean.
\newblock Efficient estimation of word representations in vector space.
\newblock In {\em 1st International Conference on Learning Representations,
  {ICLR} 2013, Scottsdale, Arizona, USA, May 2-4, 2013, Workshop Track
  Proceedings}, 2013.

\bibitem{maaten2008visualizing}
Laurens van~der Maaten and Geoffrey Hinton.
\newblock Visualizing data using t-sne.
\newblock {\em Journal of machine learning research}, 9(Nov):2579--2605, 2008.

\bibitem{rousseeuw1987silhouettes}
Peter~J Rousseeuw.
\newblock Silhouettes: a graphical aid to the interpretation and validation of
  cluster analysis.
\newblock {\em Journal of computational and applied mathematics}, 20:53--65,
  1987.

\bibitem{kaya2012sentiment}
Mesut Kaya, Guven Fidan, and Ismail~H Toroslu.
\newblock Sentiment analysis of turkish political news.
\newblock In {\em Proceedings of the The 2012 IEEE/WIC/ACM International Joint
  Conferences on Web Intelligence and Intelligent Agent Technology-Volume 01},
  pages 174--180. IEEE Computer Society, 2012.

\bibitem{ozturk2018sentiment}
Nazan {\"O}zt{\"u}rk and Serkan Ayvaz.
\newblock Sentiment analysis on twitter: A text mining approach to the syrian
  refugee crisis.
\newblock {\em Telematics and Informatics}, 35(1):136--147, 2018.

\bibitem{radford2017learning}
Alec Radford, Rafal Jozefowicz, and Ilya Sutskever.
\newblock Learning to generate reviews and discovering sentiment.
\newblock {\em arXiv preprint arXiv:1704.01444}, 2017.

\bibitem{manning-EtAl:2014:P14-5}
Christopher~D. Manning, Mihai Surdeanu, John Bauer, Jenny Finkel, Steven~J.
  Bethard, and David McClosky.
\newblock The {Stanford} {CoreNLP} natural language processing toolkit.
\newblock In {\em Association for Computational Linguistics (ACL) System
  Demonstrations}, pages 55--60, 2014.

\bibitem{taboada2011lexicon}
Maite Taboada, Julian Brooke, Milan Tofiloski, Kimberly Voll, and Manfred
  Stede.
\newblock Lexicon-based methods for sentiment analysis.
\newblock {\em Computational linguistics}, 37(2):267--307, 2011.

\bibitem{velikovich2010viability}
Leonid Velikovich, Sasha Blair-Goldensohn, Kerry Hannan, and Ryan McDonald.
\newblock The viability of web-derived polarity lexicons.
\newblock In {\em NAACL-HLT}, pages 777--785. Association for Computational
  Linguistics, 2010.

\bibitem{hamilton2016inducing}
William~L Hamilton, Kevin Clark, Jure Leskovec, and Dan Jurafsky.
\newblock Inducing domain-specific sentiment lexicons from unlabeled corpora.
\newblock In {\em Proceedings of EMNLP}, volume 2016, page 595. NIH Public
  Access, 2016.

\bibitem{o2010tweets}
Brendan O'Connor, Ramnath Balasubramanyan, Bryan~R Routledge, and Noah~A Smith.
\newblock From tweets to polls: Linking text sentiment to public opinion time
  series.
\newblock In {\em Fourth International AAAI Conference on Weblogs and Social
  Media}, 2010.

\bibitem{kashmir}
Shriphani Palakodety, Ashiqur~R. KhudaBukhsh, and Jaime~G. Carbonell.
\newblock Kashmir: {A} computational analysis of the voice of peace.
\newblock {\em CoRR}, abs/1909.12940, 2019.

\bibitem{del2017hate}
Fabio Del~Vigna, Andrea Cimino, Felice Dell’Orletta, Marinella Petrocchi, and
  Maurizio Tesconi.
\newblock {Hate me, hate me not: Hate speech detection on Facebook}.
\newblock {\em Proceedings of the First Italian Conference on Cybersecurity},
  2017.

\bibitem{davidson2017automated}
Thomas Davidson, Dana Warmsley, Michael Macy, and Ingmar Weber.
\newblock Automated hate speech detection and the problem of offensive
  language.
\newblock In {\em Eleventh International AAAI Conference on Web and Social
  Media}, 2017.

\bibitem{chandrasekharan2017you}
Eshwar Chandrasekharan, Umashanthi Pavalanathan, Anirudh Srinivasan, Adam
  Glynn, Jacob Eisenstein, and Eric Gilbert.
\newblock {You Can't Stay Here: The Efficacy of Reddit's 2015 Ban Examined
  Through Hate Speech}.
\newblock {\em Proceedings of the ACM on Human-Computer Interaction},
  1(CSCW):31, 2017.

\bibitem{dinakar2012common}
Karthik Dinakar, Birago Jones, Catherine Havasi, Henry Lieberman, and Rosalind
  Picard.
\newblock Common sense reasoning for detection, prevention, and mitigation of
  cyberbullying.
\newblock {\em ACM Transactions on Interactive Intelligent Systems (TiiS)},
  2(3):18, 2012.

\bibitem{liu2018forecasting}
Ping Liu, Joshua Guberman, Libby Hemphill, and Aron Culotta.
\newblock Forecasting the presence and intensity of hostility on instagram
  using linguistic and social features.
\newblock In {\em Twelfth International AAAI Conference on Web and Social
  Media}, 2018.

\bibitem{sindhwani2009uncertainty}
Vikas Sindhwani, Prem Melville, and Richard~D Lawrence.
\newblock Uncertainty sampling and transductive experimental design for active
  dual supervision.
\newblock In {\em Proceedings of the 26th ICML}, pages 953--960. ACM, 2009.

\bibitem{manning1999foundations}
Christopher~D Manning, Christopher~D Manning, and Hinrich Sch{\"u}tze.
\newblock {\em Foundations of statistical natural language processing}.
\newblock MIT press, 1999.

\bibitem{carneiro2009google}
Herman~Anthony Carneiro and Eleftherios Mylonakis.
\newblock Google trends: a web-based tool for real-time surveillance of disease
  outbreaks.
\newblock {\em Clinical infectious diseases}, 49(10):1557--1564, 2009.

\bibitem{koehn2005europarl}
Philipp Koehn.
\newblock Europarl: A parallel corpus for statistical machine translation.
\newblock In {\em MT summit}, volume~5, pages 79--86, 2005.

\bibitem{tiedemann2012parallel}
J{\"o}rg Tiedemann.
\newblock Parallel data, tools and interfaces in opus.
\newblock In {\em Lrec}, volume 2012, pages 2214--2218, 2012.

\bibitem{kapoor2018mind}
Raghav Kapoor, Yaman Kumar, Kshitij Rajput, Rajiv~Ratn Shah, Ponnurangam
  Kumaraguru, and Roger Zimmermann.
\newblock Mind your language: Abuse and offense detection for code-switched
  languages.
\newblock {\em arXiv preprint arXiv:1809.08652}, 2018.

\bibitem{bohra2018dataset}
Aditya Bohra, Deepanshu Vijay, Vinay Singh, Syed~Sarfaraz Akhtar, and Manish
  Shrivastava.
\newblock A dataset of hindi-english code-mixed social media text for hate
  speech detection.
\newblock In {\em Second Workshop on Computational Modeling of People’s
  Opinions, Personality, and Emotions in Social Media}, pages 36--41, 2018.

\bibitem{rubin1994social}
Jeffrey~Z Rubin, Dean~G Pruitt, and Sung~Hee Kim.
\newblock {\em Social conflict: Escalation, stalemate, and settlement}.
\newblock Mcgraw-Hill Book Company, 1994.

\end{thebibliography}
\newpage

\section{Appendix}~\label{sec:appendix}

\subsection{A Brief History of the Conflict}

 Kashmir has been a disputed region for an almost century-long India-Pakistan conflict. A key factor for continual unrest in South-East Asia, the Kashmir issue has drawn wide attention from the political science community for decades~\cite{malik2002kashmir, schofield2010kashmir, bose2009kashmir}. The root of this conflict can be traced back to the independence struggle of India and the subsequent partition of India and Pakistan in 1947. Overall, India and Pakistan have gone to full-fledged war for four times (1947, 1965, 1971 and 1999) of which, the 1971 war was the goriest one (11,000 killed from both sides) which resulted in the largest number of prisoners of war (90,000 POWs) since the Second World War~\cite{ali1983can}. In the four wars, overall, an estimated 27,650 soldiers got killed and thousands got wounded. A timeline outlining some of the key events in the bilateral conflict lasting decades is presented  below\footnote{\url{https://www.bbc.com/news/world-south-asia-16069078}}.  

\vspace{0.5cm}

\scalebox{1}{

\begin{tabular}{r |@{\foo} p{0.85\textwidth}}

1947 & British rule over Indian subcontinent ends. Territory is partitioned into
the Islamic Republic of Pakistan (Muslim majority), and the Republic of India (Hindu majority). Kashmir (Muslim majority) becomes a sovereign monarchy led by a
Hindu dynasty.\\
1947 & Pakistani tribal military attacks Kashmir. Merely two months after declaring independence from both India and Pakistan, the ruler of Kashmir signs a treaty of accession to India triggering the first war between India and Pakistan. \\
1948 & The Kashmir issue is discussed at the United Nations (UN) Security Council. Ceasefire is achieved in 1949 but Pakistan does not evacuate its troops thereby partitioning Kashmir. The modern territories of Pakistan-administered Kashmir - Azad Kashmir and Gilgit-Baltistan (then Northern Areas), and India-administered Jammu and Kashmir are formed. The UN resolution calls for a referendum on the status of the Kashmir region.\\
1951 & Elections in India-administered Jammu and Kashmir show strong support for Indian accession and India considers a referendum unnecessary. Pakistan and the UN disagree on
the counts that the Pakistan-administered regions were not considered in the vote.\\
1965 & India and Pakistan go to war for the second time over Kashmir. The brief war ends in a ceasefire and a return to the original positions.\\
1971-72 & India and Pakistan go to war for the third time. The war results in a defeat for Pakistan and the Simla Agreement is signed. The Kashmir ceasefire line is christened the Line of Control (LoC). Both sides pledge to settle the Kashmir issue through negotiations.\\
1984 & The Siachen glacier, a vital strategic territorial asset, is seized by the Indian military. The glacier heretofore not demarcated by the LoC is the casus belli of several future confrontations between India and Pakistan. The Siachen glacier is the highest warzone in the world.\\
1998 & India's Prime Minister declares India a full-fledged nuclear state in a press conference. Shortly after, Pakistan successfully develops and tests its own nuclear weapons and both nations become one of a handful of global nuclear powers.\\
1999 & India and Pakistan go to war for the fourth time. The event was triggered by
militant activity in the Indian-administered Kargil district of Jammu and Kashmir. The
conflict (called the Kargil war) lasts for approximately 2 months.\\
2000-19 & In a flurry of terror attacks in both nations where both parties allege the involvement of the other, \texttt{PULWAMA} marks the most recent episode bringing both
the nuclear states precariously close to war.
\end{tabular}

}

\subsection{Data set: YouTube comments}

 We obtain a comprehensive set of YouTube comments using the publicly available YouTube API on incident-specific videos during our period of interest (14th Feb 2019 to 13th Mar 2019). In what follows, we first provide a brief outline of our data collection process and then describe each of the steps in greater detail. 

Our data collection procedure consists of the following steps:
\begin{enumerate}
\item Start with a small seed set, $\mathcal{S}$, with potentially relevant search queries. 
\item Expand $\mathcal{S}$ to $\mathcal{S}_{\emph{related}}$ by including related search queries from Google Trends\footnote{\url{https://trends.google.com/trends/?geo=US}} during the period of interest. 
\item Create a set of queries, $\emph{News}$, containing ten popular world news channels, and ten popular news channels from India and Pakistan. Expand $\mathcal{S}_{\emph{related}}$ to $\mathcal{S}_{\emph{news}}$ which contains all unique queries in $\mathcal{S}_{\emph{related}}$ and $\mathcal{S}_{\emph{related}} \times \emph{News}$.
\item Query YouTube for videos posted within the specified time range. Obtain a video set, $\mathcal{V}$, containing top 10 video search results for each query in $\mathcal{S}_{\emph{news}}$ and top 200 video recommendations for each query in $\mathcal{S}_{\emph{related}}$.
\item Manually inspect $\mathcal{V}$, remove irrelevant videos and obtain $\mathcal{V}_{\emph{relevant}}$ consisting of relevant videos. 
\item Prune $\mathcal{V}_{\emph{relevant}}$ further to set of popular videos, $\mathcal{V}'_{\emph{relevant}}$,  by removing any video that has 10 or fewer comments. 
\item Obtain $\mathcal{C}_{\emph{all}}$, the set of user comments to every video in $\mathcal{V}'_{\emph{relevant}}$.
\item Construct $\mathcal{C}_{\emph{en}}$ by filtering $\mathcal{C}_{\emph{all}}$  further to restrict it to comments written in English (described later).  
\end{enumerate}

We now provide a detailed description of our procedure. In step 1, we constructed $\mathcal{S}$ with the following six queries: \texttt{[pulwama]}, \texttt{[balakot]}, \texttt{[abhinandan]},  \texttt{[kashmir]}, \texttt{[india pakistan]}, \texttt{[india pakistan war]}.

We expanded each query in $\mathcal{S}$ with related search queries procured from Google Trends. Google Trends queries have been found effective in tackling time-series AI problems like early-detection of disease outbreaks~\cite{carneiro2009google}. Google Trends allows to specify geographic location of interest and period of interest. For each query in $\mathcal{S}$, we set the location of interest to India or Pakistan and the start date as February 14th, 2019 and the end date as March 13th, 2019. Our first four seed queries have only a single token. However, we noticed that the queries \texttt{[India Pakistan]} and \texttt{[Pakistan India]} yielded slightly different results, Following \cite{stephens2017everybody}, that revealed that we tend to put our more-preferred choice ahead in a pair, whenever we have a query that contained a country pair (e.g., \texttt{[India Pakistan]} or \texttt{[Pakistan India war]}, we adjusted the order of the pair accordingly matching it with the location of interest. In step 2, upon expansion using Google Trends and subsequent removal of duplicate queries, we obtained $\mathcal{S}_\emph{related}$ with 207 unique queries. 

\begin{table}[t]
{
\begin{center}
\begin{tabular}{ |l|c|c| } 
 \hline
    India & Pakistan  & World \\
    \hline
    Times Now & ARY News & Geo News \\
 \hline
  India Today & Geo News & NDTV India \\
 \hline
  Aaj Tak & Dunya News & Euro news \\
 \hline
  NDTV 24x7 & Samaa TV & Al Jazeera \\
 \hline
  ABP News & CNN & Al Arabiya \\
 \hline
  CNN-News 18 & 92 News & MSNBC \\
 \hline
  India TV & PTV News & Sky News \\
 \hline
  BBC World News & Hum News & CNN \\
 \hline
  NDTV India & Neo News & Fox News \\
 \hline
  Zee News & Aaj News & BBC News \\
 \hline
\end{tabular}
\end{center}
\caption{\small{List of popular news channels we considered.}}
\label{tab:channels}
}
\end{table}

In step 3, with 29 unique news channels in $\emph{News}$, the Cartesian-product between $\mathcal{S}_{\emph{related}}$ and  $\emph{News}$ (listed in Table~\ref{tab:channels}) produced 6003 queries in $\mathcal{S}_{\emph{news}}$. 

In step 4, we obtained 6,157 unique videos and after removing irrelevant and unpopular videos (less than 10 comments), we finally obtained a set of 2,890 videos. Two annotators fluent in English, Hindi and Urdu and familiarity with additional European and Indian languages annotated the videos (inter-rater agreement Cohen's $\kappa$: 0.87). We considered the consensus labels.

$\mathcal{C}_{\emph{all}}$ consists of 2.04 million user comments coming from 791,289 unique users. After removing punctuation and emojis and lower-casing the comments written in roman alphabet, the average length of a comment was 22.51 $\pm$ 38.47 tokens.

We finally ran our language identification module  and obtained $\mathcal{C_{\emph{en}}}$, a set of 921,235 YouTube comments posted by 392,460 users. 

\newpage

\subsection{Language Identification}~\label{sec:langID}

We now provide a thorough treatment of our approach on additional data sets. We focus on two low-resource language settings (Bengali and Oriya) and a large mix of well-formed texts from Europe (21 languages, EuroParl data set~\cite{koehn2005europarl}). We first make two slight digressions: one to provide an intuition for our our results, the other to address the issue of fairness in comparison. 

\noindent\textbf{Intuition:} The SkipGram model used for training the FastText embeddings predicts an input word's context. 
In a polyglot setting, the likeliest context predicted for a Hindi word is other Hindi
words. The embeddings likely reflect this aspect of the language model and
thus we see language clusters. We admit that implementation choices like
splitting on whitespace (for instance) can preclude some languages, so
we refrain from making claims about the universality of the technique and
present empirical results only on Indian and European languages.

\begin{figure}[t]
\centering
\includegraphics[trim={0 0 0 0},clip, width=3.4in, height = 2.2in]{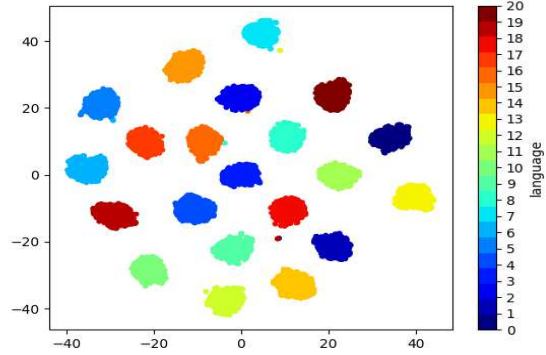}
\caption{A visualization of the polyglot document-embedding space for the $\mathcal{D}_{Euro}$ data set.}
\label{fig:cluster-euro}
\end{figure}

\noindent\textbf{Fairness:} A fair performance comparison between the two supervised baselines: \texttt{GoogleLangID} and \texttt{fastTextLangID} and our proposed unsupervised approach is challenging for the following reasons. On one hand the baselines predict from a larger set of languages. In contrast, our method reveals only those languages observed in the corpus in question - thus a limited set of clusters (labels) is obtained - in most cases this is substantially smaller than the number of languages supported by industrial strength baselines. On the other hand, the baselines are supervised methods that have been trained on a vast amount of data whereas our methods require minimal manual labeling - a critical feature for dealing
with corpora featuring low resource languages which are a common occurrence in the Indian subcontinent.

We admit that restricting the baselines to predict only from the smaller set may offset the advantage of our method. The API for \texttt{fastTextLangID} provides an ordered list of all languages that it supports with the confidence score. Let the set of all languages present in a corpus be denoted as $\mathcal{L}$. For a given document, we predict the language belonging to $\mathcal{L}$ with the highest confidence score. Suppose the top three predictions for a document from our India-Pakistan data set by \texttt{fastTextLangID} are (1) \emph{German} (predicted with highest confidence), (2) \emph{Spanish} and (3) \emph{Hindi}. Since \emph{Hindi} $\in \mathcal{L}$, and \emph{German} $\notin \mathcal{L}$, \emph{Spanish} $\notin \mathcal{L}$, we consider that the predicted label is \emph{Hindi}. We denote this new setting as \texttt{fastTextLangID}$_{\emph{fair}}$. For $\mathcal{C}_{\emph{all}}$, we present in the performance of this additional baseline in Table~\ref{tab:kashmirResultsAppendix}.

\noindent\textbf{Data sets:} We now describe our additional data sets, two of which are collected from the Indian subcontinent, one is a well-known data set of European languages. 

\begin{compactitem}
    \item $\mathcal{D}_{ABP}$: The ABP Ananda news channel is a Bengali news organization.
    We crawled the comments on videos uploaded by their YouTube channel\footnote{\url{https://www.youtube.com/channel/UCv3rFzn-GHGtqzXiaq3sWNg}}
    and obtained 219,927 comments.
    Most of the comments are in Bengali,
    Hindi, and English. Note that internet users in the Indian subcontinent use the Latin script
    as well as their native script for writing. The use of the Latin script for writing in Hindi
    and Bengali is significant in this corpus.
    \item $\mathcal{D}_{OTV}$: OTV is an Oriya news network with a popular YouTube channel\footnote{\url{https://www.youtube.com/channel/UCCgLMMp4lv7fSD2sBz1Ai6Q}}. We crawled videos
    from this network and subsequently crawled comments to obtain 153,435 comments.
    with most of the comments posted in Oriya, Hindi, and
    English. Latin script is heavily used alongside the native script for Oriya and Hindi.
    \item $\mathcal{D}_{Euro}$: The Europarl corpus~\cite{koehn2005europarl} contains 21 languages with well-written text.
    The processed version is obtained from ~\cite{tiedemann2012parallel}. 420,000 documents were
    reserved for training and 210,000 documents were used for test.
\end{compactitem}

\begin{table}[t]
  \centering
  \begin{tabular}{|l|c|l|c|c|c|}
    \hline
    Method & Accuracy & Language & P & R & F1\\
    \hline
    Our method & \textbf{0.99} & Hindi (E) (52.5\%)  & \textbf{1.0} & \textbf{0.98}  & \textbf{0.99} \\
               &      & English  (46.5\%) & \textbf{0.99}  & \textbf{1.0}  & \textbf{0.99} \\
               &      & Hindi (1\%)  & \textbf{1.0} & \textbf{1.0}  & \textbf{1.0} \\
    \hline
    \texttt{fastTextLangID} & 0.48 & Hindi (E) (52.5\%)   & \textbf{1.0}   & 0.01  & 0.02 \\
             &       & English  (46.5\%) & 0.55  & \textbf{1.0}  & 0.71 \\
             &       & Hindi (1\%)  & \textbf{1.0}   & \textbf{1.0}   & \textbf{1.0} \\
    \hline
    \texttt{fastTextLangID}$_{\emph{fair}}$ & 0.48 & Hindi (E) (52.5\%)   & \textbf{1.0}   & 0.01  & 0.02 \\
             &       & English  (46.5\%) & 0.55  & \textbf{1.0}  & 0.71 \\
             &       & Hindi (1\%)  & \textbf{1.0}   & \textbf{1.0}   & \textbf{1.0} \\
    \hline
    
    \texttt{GoogleLangID} & 0.96 & Hindi (E) (52.5\%)   & 0.97 & 0.94 & 0.96 \\
                 &      & English  (46.5\%) & 0.97 & 0.97 & 0.97 \\
                 &      & Hindi (1\%)  & 0.4 & \textbf{1.0}  & 0.57 \\
    \hline
    \end{tabular}
\vspace{0.5cm}
   \caption{$\mathcal{C}_{\emph{all}}$. Language written in Roman script is indicated with (E). 
    Percentage of the ground truth assigned this label is indicated for each language. Best metric is highlighted in bold for
    each language. P: precision, R: recall.}
    \label{tab:kashmirResultsAppendix}
\end{table}

\begin{table}[htb]
  \centering
  \begin{tabular}{|l|l|c|c|c|c|}
    \hline
    Method & Accuracy & Language & P & R & F1  \\
    \hline
    Our Method & \textbf{0.985} & Oriya (E) (65.5\%) & \textbf{1.0}  & \textbf{0.98} & \textbf{0.99} \\
               &       & Oriya (6.5\%)  & \textbf{1.0}  & \textbf{1.0}  & \textbf{1.0} \\
               &       & English  (18.5\%) & \textbf{1.0}  & \textbf{1.0}  & \textbf{1.0} \\
               &       & Hindi (E) (9.5\%)  & \textbf{0.86} & \textbf{1.0}  & \textbf{0.93} \\
    \hline
    \texttt{fastTextLangID} & 0.25 & Oriya (E) (65.5\%) & 0.0  & 0.0 & 0.0 \\
             &       & Oriya (6.5\%)  & \textbf{1.0}  & \textbf{1.0}  & \textbf{1.0} \\
             &       & English  (18.5\%) & 0.23  & 1.0  & 0.38 \\
             &       & Hindi (E) (9.5\%)  & 0.0 & 0.0  & 0.0 \\
    \hline
    \texttt{fastTextLangID}$_{\emph{fair}}$ & 0.25 & Oriya (E) (65.5\%) & 0 & 0.0 & 0.0\\
             &       & Oriya (6.5\%)  & \textbf{1.0} & \textbf{1.0} & \textbf{1.0} \\
             &       & English  (18.5\%) & 0.19& 1.0& 0.33 \\
             &       & Hindi (E) (9.5\%)  & 0 & 0 & 0 \\
    \hline
    \texttt{GoogleLangId} & 0.26 & Oriya (E) (65.5\%) & 0.0  & 0.0  & 0.0 \\
                 &      & Oriya (6.5\%)  & 0.0  & 0.0  & 0.0 \\
                 &      & English  (18.5\%)     & 0.92 & 0.97 & 0.95 \\
                 &      & Hindi (E) (9.5\%)  & 0.38 & 0.84 & 0.52 \\
    \hline
    \end{tabular}
    \vspace{0.5cm}
    \caption{{Performance evaluation on $\mathcal{D}_{\emph{OTV}}$}. Language written in Latin script is indicated with (E). 
    Percentage of the ground truth assigned this label is indicated for each language. Best metric is highlighted in bold for
    each language.}
  \label{tab:otv}
\end{table}

\begin{table}[htb]
  \centering
  \begin{tabular}{|l|c|l|c|c|c|}
    \hline
    Method & Accuracy & Language & P & R & F1  \\
    \hline
    Our Method & \textbf{0.96} & Bengali(E) (54\%) & \textbf{1.0} & \textbf{0.95} & \textbf{0.98} \\
               &      & Bengali (22.5\%) & \textbf{1.0} & \textbf{1.0} & \textbf{1.0} \\
               &      & English (18\%) & \textbf{1.0} & 0.92 & 0.96 \\
               &      & Hindi(E) (5\%) & \textbf{0.53} & \textbf{1.0} & \textbf{0.69} \\
               &      & Hindi (0.5\%)  & 0.0 & 0.0  & 0.0 \\
    \hline
    \texttt{fastTextLangID} & 0.4 & Bengali(E) (54\%) & 0 & 0 & 0 \\
             &      & Bengali (22.5\%) & \textbf{1.0} & \textbf{1.0} & \textbf{1.0} \\
             &      & English  (18\%) & 0.34 & 0.94 & 0.50 \\
             &      & Hindi(E) (5\%) & 0 & 0.0 & 0.0 \\
             &      & Hindi (0.5\%) & \textbf{1.0} & \textbf{1.0} & \textbf{1.0} \\
    \hline
        \texttt{fastTextLangID}$_{\emph{fair}}$ & 0.42 & Bengali(E) (54\%) & \textbf{1.0} & 0.02 & 0.04  \\
             &      & Bengali (22.5\%) & \textbf{1.0} & \textbf{1.0} & \textbf{1.0} \\
             &      & English  (18\%) & 0.24 & \textbf{1.0} & 0.39 \\
             &      & Hindi(E) (5\%) & 0 & 0.0 & 0.0 \\
             &      & Hindi (0.5\%) & 0.5 & \textbf{1.0} & 0.66 \\
    \hline

    \texttt{GoogleLangId} & 0.91 & Bengali(E) (54\%) & 0.99 & 0.87 & 0.93 \\
                 &      & Bengali (22.5\%) & 0.98 & \textbf{1.0} & 0.99 \\
                 &      & English  (18\%) & 0.97 & \textbf{0.97} & \textbf{0.97} \\
                 &      & Hindi(E) (5\%) & 0.5 & 0.7 & 0.58 \\
                 &      & Hindi (0.5\%) & 0.13 & \textbf{1.0} & 0.22 \\
    \hline
    \end{tabular}
    \vspace{0.5cm}
    \caption{{$\mathcal{D}_{\emph{ABP}}$}. Language written in Latin script is indicated with (E). 
    Percentage of the ground truth assigned this label is indicated for each language. Best metric is highlighted in bold for
    each language.}
  \label{tab:abp}
\end{table}

\noindent\textbf{Performance on EuroParl:} Our model's performance is on-par with \verb|fastTextLangID|. We did not evaluate against \verb|GoogleLangId| due to prohibitive costs and it is reasonable to expect very high accuracy due to the clean nature of the corpus. Our method is near-perfect and on-par with \verb|fastTextLangID|. Our model's accuracy is $99.9\%$ versus $99.3\%$ for \verb|fastTextLangID|. 

\noindent\textbf{Performance on low resource languages:} We considered two additional data sets consisting of a mix of languages of which two are low resource languages (Bengali and Oriya). As shown in Table~\ref{tab:abp} and \ref{tab:otv}, our performance on the India-Pakistan data set translates to other languages in the Indian subcontinent. We observed that our added fairness criterion marginally improved the performance of \texttt{fastTextLangID} but our method still substantially outperformed \texttt{fastTextLangID}$_{\emph{fair}}$. As we already mentioned, we could not construct a similar \texttt{GoogleLangID}$_{\emph{fair}}$ due to its API's limitation. However, based on our current observations on \texttt{fastTextLangID}$_{\emph{fair}}$, we conjecture that the performance boost would not be substantial.

\end{document}